\documentclass[11pt]{article}

\usepackage{amsmath}

\begin{document}


\newcommand{\N}{N\raise.7ex\hbox{\underline{$\circ $}}$\;$}




\title{
  V.V. Kisel, E.M. Ovsiyuk, V.M. Red'kov\\[3mm]
Particle with spin 2 and anomalous magnetic moment \\
in external
electromagnetic and gravitational fields }

\maketitle

\begin{quotation}

Tensor 50-component form of the  first order relativistic wave equation for a particle with spin 2 and anomalous magnetic
moment is extended to the case of an arbitrary   curved space -- time geometry.
An additional parameter considered in  the presence of only electromagnetic field as related to
anomalous magnetic  moment,
turns to determine  additional interaction terms with external  geometrical background through
Ricci  $R_{kl}$ and Riemann  $R_{klmn}$ tensors.

\end{quotation}

Theory of massive and massless fields of spin 2, staring from
Pauli and Fierz investigations  \cite{1939-Pauli-Fierz, 1939-Fierz-Pauli},
always were attracted attention: de Broglie  \cite{1941-Broglie},
Pauli  \cite{1941-Pauli}, Gel'fand--Yaglom
\cite{1948-Gel'fand-Yaglom}, Fradkin  \cite{1950-Fradkin},
 Fedorove et al . \cite{1951-Fedorov, 1967-Krylov,
 1968-Bogush,  1972-Krylov, 1986-Kisel},
 Fainberg  \cite{1955-Fainberg},
 Regge \cite{1957-Regge},
  Buchdah  \cite{1958-Buchdahl, 1962-Buchdahl},
Velo -- Zwanziger \cite{1969-Velo-Zwanziger, 1972-Velo},
 Cox  \cite{1982-Cox}, Barut \cite{1982-Barut},
 Loide  \cite{1986-Loide}, Vasiliev
\cite{1992-Vasiliev}, Buchbinder et al . \cite{1999-Buchbinder,
2000-Buchbinder(1)}, Bogush et al \cite{20003-Bogush-Kisel-Tokarevskaya-Red'kov, 20030-Red'kov-Tokarevskaya-Kisel,
2010-Kisel-REd'kov(1), 2010-Kisel-REd'kov(2)}).
Most of investigations were performed within second order wave equations approach.
However, it is known that many of problems arising for fields of higher
 spin may be avoided if one   starts with the theory of first order equations.
One of the earliest considerations of the theory for spin 2 particle was given by Fedorov
 \cite{1951-Fedorov}. It turned out that such a description requires
 30 component wave function. Afterwards
Fedorov proposed else one description for spin 2 particle, 50-component model,
 \cite{1968-Fedorov, 1976-Fedorov-Kirilov}. The primary question is about  relation between two models.
Bogush and Kisel   \cite{1988-Bogush-Kisel} demonstrated  (within spinor formalism)
that  50-component model describes a spin 2 particle with
additional electromagnetic characteristics, anomalous magnetic moment\footnote{Here one may see analogy with the known Petras  \cite{1955-Petras(1),
 1955-Petras(2), 1955-Formanek, 1984-Bogush(2)} theory for a spin $1/2$ particle, or
Shamali -- Capri
  \cite{1969-Capri(1), 1969-Capri(2), 1971-Shamali, 1972-Capri, 1973-Shamaly, 1974-Hagen}
  theory for a particle with spin 1, when  increase in the number of field variables
permits us to introduce an additional parameter for a particle, anomalous magnetic  moment.
}.
More detailed analysis of this spinor description was given in
 \cite{1984-Bogush(1), 1984-Bogush(3),1984-Bogush(3), 1984-Kisel, 1988-Bogush-Kisel}.
In  \cite{2010-Kisel-REd'kov(1), 2010-Kisel-REd'kov(2)}, that theory was transformed to
more a simple tensor  technique.

In the present paper we consider a 50-component model for a massive spin 2 particle
in presence of  external electromagnetic and gravitational fields. The primary question is how additional
intrinsic structure of the particle manifests itself in any curved space-time background.

We start with  tensor equations given  in \cite{2010-Kisel-REd'kov(1), 2010-Kisel-REd'kov(2)}
for  flat Minkowski space-time, and extend them by changing ordinary derivative into covariant ones
( $\partial_{b} \Longrightarrow \nabla_{\beta}$), so we arrive at
$$
2\;  \lambda_{1} D^{a} \Psi_{a}^{(1)} +  2\; \lambda_{2} D^{a}
\Psi_{a}^{(2)} \; ) + i M\; \Psi = 0 \; , \eqno(1a)
$$

$$
\lambda _{3} \;  D_{a} \Psi + 2  \lambda_{4} \; D^{b} \Psi_{(ba)}+
iM\; \Psi^{(1)}_{a} =0 \; , \eqno(1b)
$$
$$
\lambda _{5} \;  D_{a} \Psi + 2  \lambda_{6} \; D^{b} \Psi_{(ba)}+
iM\; \Psi^{(2)}_{a} =0 \; , \eqno(1c)
$$
$$
{ \lambda_{7} \over 2}  \;  \left ( D_{a} \Psi^{(1)}_{b}  + D_{b}
\Psi^{(1)}_{a} -{1 \over 2}  \; g_{ab} D^{c} \Psi^{(1)}_{c} \right
)
$$
$$
+
 { \lambda_{8} \over 2}  \;  \left ( D_{a} \Psi^{(2)}_{b}  +
D_{b} \Psi^{(2)}_{a} -{1 \over 2}  \; g_{ab} D^{c} \Psi^{(2)}_{c}
\right ) +
$$
$$
+ 2 \lambda_{9} \; D^{c} \Psi_{(abc)} -2\lambda_{10} \; \left  (
\; D^{c} \Psi_{a[bc]} + D^{c} \Psi_{b[ac]}  \; \right ) + i M\;
\Psi_{(ab)} =0  \; ,
$$
$$
\eqno(1d)
$$
$$
{\lambda_{11} \over 2}\; \left (  \; D_{c} \Psi_{(ab)} - D_{b}
\Psi_{(ac)} -
 {1 \over 3} g_{ca} D^{m} \Psi_{(mb)} +
{1 \over 3} g_{ba}  D^{m} \Psi_{(mc)}  \; \right ) + i M \;
\Psi_{a[bc]} = 0 \; ,
$$
$$
\eqno(1e)
$$
$$
{\lambda_{12} \over 3} \; \left  (\;  D_{a} \Psi_{(bc)} + D_{b}
\Psi_{(ca)} +
 D_{c} \Psi_{(ab)}  \right.
 $$
 $$
\left. - {1\over 3}\; g_{ac} D^{m}\Psi_{(mb)} - {1\over 3}\;
g_{cb} D^{m}\Psi_{(ma)} - {1\over 3}\; g_{ba} D^{m}\Psi_{(mc)} \;
\right  ) + i M \Psi_{(abc)}=0 \; .
 $$
 $$
 \eqno(1f)
 $$

\noindent Here $ D_{a} =  \nabla_{a} +ie A_{a}$,  where  $\nabla_{a}$ is a covariant derivative,
$A_{a}$ stands for electromagnetic potential;
$\lambda_{1}, ...,\lambda_{12} $ are 12 numerical constants obeying
additional restrictions\footnote{Below it will be clear that only one parameter
has physical sense, referring to anomalous magnetic  moment,  all other  can be eliminated  from the model.}
$$
2\lambda_{10} \lambda_{11} -{2 \over 3} \lambda_{9} \lambda_{12} = 1 \;,
\qquad
\lambda_{4}\lambda_{7}  + \lambda_{6}\lambda_{8}  + {8\over 9} \lambda_{9}\lambda_{12} = {1 \over 3}\; ,
$$
$$
\lambda_{1} \lambda_{3} +  \lambda_{2} \lambda_{5} = -{1 \over 4} \;,
\qquad
(\lambda_{1} \lambda_{4} +  \lambda_{2} \lambda_{6}) \;
(\lambda_{3} \lambda_{7} +  \lambda_{5} \lambda_{8})  = -{1 \over 12} \; .
\eqno(2a)
$$

\noindent
In  50-component model for   a spin 2 particle, we employ
one scalar, two vectors, and three tensors:
$$
\Psi \;  , \qquad \Psi^{(1)}_{a} \;, \;  \Psi^{(2)}_{a}\; ,\qquad \Psi_{(ab)}\; , \; \Psi_{a[bc]} \; , \;  \Psi_{(abc)} \; .
\eqno(2b)
$$

 Recall that in 30-component model there involved scalar, vector, and two tensors
 $$
\Phi \;  , \qquad \Phi_{a} \;,  \qquad \Phi_{(ab)}\; , \; \Phi_{[ab]c} \; ;
\eqno(3a)
$$
with 30 independent variables
$$
\Phi (x) \;\; \Longrightarrow \;\; 1 \; , \qquad
\Phi_{a}  \; \;  \Longrightarrow \;\; 4 \; ,
$$
$$
\Phi_{(ab)}  \;\; \Longrightarrow \;\; (10 - 1) = 9 \; , \qquad
\Phi_{[ab]c} \;\; \Longrightarrow \;\; 6 \times 4 - 4 -4  = 16 \; ;
$$
\noindent and  equations  (compare with  (1a)--(1f))
$$
D^{a} \Phi _{a} -M  \; \Phi = 0 \; , \qquad
{1 \over 2} \; D_{a} \Phi  -  {1 \over 3} \;
D^{b} \Phi_{(ab)} -M \; \Phi_{a} = 0 \; ,
$$
$$
D_{a} \Phi_{b} \; +\;  D_{b} \Phi_{a} -
{1 \over 2} g_{ab} \;  D^{k} \Phi_{k}
+{1 \over 2} \; ( D^{k} \Phi_{[ka]b}
$$
$$
 +  D^{k}
\Phi_{[kb]a} -   {1 \over 2} g_{ab} D^{k}
\Phi_{[kn]}^{\;\;\;\;n}\; )  - M \; \Phi_{(ab)}  = 0\;
,
$$
$$
D_{a} \Phi_{(bc)}  -  D_{b} \Phi_{(ac)}  +   {1 \over 3} \;
(g_{bc}  D^{k} \Phi_{(ak)}  -   g_{ac} \; D^{k}
\Phi_{(bk)} )  - M \; \Phi _{[ab]c}  = 0\; .
\eqno(3b)
$$

Below we will show that excluding from the 50-component models
superfluous variables (formally they consist a 4-vector  and 3-rank tensor)
and  introducing new field  variables, one can get a    30-component model
modified by presence  additional interaction terms  with electromagnetic and gravitational
fields.

 To this end, first instead of
  $\Psi^{(1)}_{a},  \Psi^{(2)}_{a}$  let us introduce new variables
  $$
\left | \begin{array}{c}
B_{a} \\ C_{a}
\end{array} \right | =
\left | \begin{array}{rr}
\lambda_{1} & \lambda_{2} \\
\lambda_{7} &  \lambda_{8}
\end{array} \right |
\left | \begin{array}{c}
\Psi^{(1)}_{a} \\ \Psi^{(2)}_{a}
\end{array}  \right | \; , \qquad
\left | \begin{array}{c}
\Psi^{(1)}_{a} \\ \Psi^{(2)}_{a}
\end{array}  \right |= { 1 \over  \lambda_{1} \lambda_{8} - \lambda_{2} \lambda_{7}}
\left | \begin{array}{rr}
\lambda_{8} & -\lambda_{2} \\
-\lambda_{7} &  \lambda_{1}
\end{array} \right |
\left | \begin{array}{c}
B_{a} \\ C_{a}
\end{array} \right | \; .
$$

The system (1) can be presented as follows

$$
2\;  D^{a} B _{a} + i m\; \Psi = 0 \; , \eqno(4a)
$$
$$
-{1 \over 4}  \;  D_{a} \Psi + 2  \left  (  \lambda_{1} \lambda_{4} +
\lambda_{2} \lambda_{6} \right ) \; D^{b} \Psi_{(ba)}+ iM\; B_{a} =0 \; ,
\eqno(4b)
$$
$$
(\lambda_{7} \lambda _{3} + \lambda_{8} \lambda_{5}) \;  D_{a}
\Psi + 2 \left  (  \lambda_{7} \lambda_{4} + \lambda_{8} \lambda_{6} \right ) \;
D^{b} \Psi_{(ba)}+ iM\; C_{a} =0 \; , \eqno(4c)
$$
$$
{ 1  \over 2}  \;  \left ( D_{a} C_{b}  + D_{b} C_{a} -{1 \over 2} \;
g_{ab} D^{c} C_{c} \right  ) +
$$
$$
+ 2 \lambda_{9} \; D^{c} \Psi_{(abc)} -2\lambda_{10} \;  \left ( \; D^{c}
\Psi_{a[bc]} + D^{c} \Psi_{b[ac]}  \right ) + i M\; \Psi_{(ab)} =0 \; ,
$$
$$
\eqno(4d)
$$
$$
{\lambda_{11} \over 2}\; \left ( D_{c} \Psi_{(ab)} - D_{b} \Psi_{(ac)} -
 {1 \over 3} g_{ca} D^{m} \Psi_{(mb)} +
{1 \over 3} g_{ba}  D^{m} \Psi_{(mc)} \right  ) + i M \; \Psi_{a[bc]} = 0 \; ,
$$
$$
\eqno(4e)
$$
$$
{\lambda_{12} \over 3}  \left [   D_{a} \Psi_{(bc)} +   D_{b}
\Psi_{(ca)} +
 D_{c} \Psi_{(ab)}  \right.
 $$
 $$
 \left.
 - {1\over 3}  g_{ac} D^{m}\Psi_{(mb)} - {1\over 3}\;
g_{cb} D^{m}\Psi_{(ma)} - {1\over 3} g_{ba} D^{m}\Psi_{(mc)}
\right ]  +
 iM \Psi_{(abc)}=0 \; .
$$
$$
\eqno(4f)
 $$

Multiplying eq. (4c) by
$(\lambda_{1} \lambda_{4} +  \lambda_{2} \lambda_{6})$ and taking into account  (2a), we get
$$
-{1 \over 12}
  D_{a}
\Psi + 2  \left  ( \lambda_{1} \lambda_{4} +  \lambda_{2} \lambda_{6} \right )
(  \lambda_{7} \lambda_{4} + \lambda_{8} \lambda_{6}) \;
D^{b} \Psi_{(ba)} + iM\;  \left (\lambda_{1} \lambda_{4} +  \lambda_{2} \lambda_{6} \right ) \;  C_{a} =0 \; .
$$

\noindent Substituting expression for $D_{a} \Psi$ from  (4b), we arrive at
$$
-{2 \over 3}   (  \lambda_{1} \lambda_{4} +
\lambda_{2} \lambda_{6}) \; D^{b} \Psi_{(ba)} - {iM \over 3 } \; B_{a}
        $$
    $$
    + 2   (\lambda_{1} \lambda_{4} +  \lambda_{2} \lambda_{6})
(  \lambda_{7} \lambda_{4} + \lambda_{8} \lambda_{6}) \;
D^{b} \Psi_{(ba)}+
 iM\; (\lambda_{1} \lambda_{4} +  \lambda_{2} \lambda_{6})  C_{a} =0 \; ,
$$

\noindent
from whence it follows
$$
  C_{a}  =  {1  \over 3 (  \lambda_{1} \lambda_{4} +\lambda_{2} \lambda_{6})   } \; B_{a}
 -  {2  \over  iM} \;  \left  [     (  \lambda_{7} \lambda_{4} + \lambda_{8} \lambda_{6})  - {1 \over 3}   \right  ]
D^{n} \Psi_{(na)}  \; .
 \eqno(4c')
$$

\noindent This identity permits to exclude a  superfluous vector
$C_{a}$. In particular, then eq. (4d) gives
$$
{1  \over 6 (  \lambda_{1} \lambda_{4} +\lambda_{2} \lambda_{6})   }
  \; ( D_{a} B_{b}  + D_{b} B_{a} -{1 \over 2} \; g_{ab} D^{c} B_{c} )
$$
$$
- {1  \over  iM} \;
\left  [ (  \lambda_{7} \lambda_{4} + \lambda_{8} \lambda_{6})  - {1 \over 3}   \right  ] \;
\left ( D_{a} D^{n} \Psi_{(nb)}  + D_{b} D^{n} \Psi_{(na)}  -
{1 \over 2} \; g_{ab} D^{c} D^{n} \Psi_{(nc)}  \right )
$$
$$
+ 2 \lambda_{9} \; D^{c} \Psi_{(abc)} -2\lambda_{10} \; ( \; D^{c}
\Psi_{a[bc]} + D^{c} \Psi_{b[ac]} ) + i M\; \Psi_{(ab)} =0 \; .
$$
$$
\eqno(4d')
$$

Therefore, instead of eqs. (4) we can us an equivalent one
$$
2\;  D^{a} B _{a} + i M \; \Psi = 0 \; , \eqno(4a)
$$

$$
-{1 \over 4}  \;  D_{a} \Psi + 2  (  \lambda_{1} \lambda_{4} +
\lambda_{2} \lambda_{6}) \; D^{b} \Psi_{(ba)}+ iM\; B_{a} =0 \; ,
\eqno(4b)
$$
$$
C_{a} =
 {1  \over 3 (  \lambda_{1} \lambda_{4} +\lambda_{2} \lambda_{6})   } \; B_{a}
 -  {2  \over  iM} \;  \left  [     (  \lambda_{7} \lambda_{4} + \lambda_{8} \lambda_{6})  - {1 \over 3}   \right  ]
D^{n} \Psi_{(na)}  \; ,
 $$
 $$
 \eqno(4c')
$$
$$
{1  \over 6 (  \lambda_{1} \lambda_{4} +\lambda_{2} \lambda_{6})   }
  \; ( D_{a} B_{b}  + D_{b} B_{a} -{1 \over 2} \; g_{ab} D^{c} B_{c} )
$$
$$
- {1  \over  iM} \;
\left  [ (  \lambda_{7} \lambda_{4} + \lambda_{8} \lambda_{6})  - {1 \over 3}   \right  ] \;
\left ( D_{a} D^{n} \Psi_{(nb)}  + D_{b} D^{n} \Psi_{(na)}  -
{1 \over 2} \; g_{ab} D^{c} D^{n} \Psi_{(nc)}  \right )
$$
$$
+ 2 \lambda_{9} \; D^{c} \Psi_{(abc)} -2\lambda_{10} \; ( \; D^{c}
\Psi_{a[bc]} + D^{c} \Psi_{b[ac]} ) + i M\; \Psi_{(ab)} =0 \; ,
$$
$$
\eqno(4d')
$$
$$
{\lambda_{11} \over 2}\; ( D_{c} \Psi_{(ab)} - D_{b} \Psi_{(ac)} -
 {1 \over 3} g_{ca} D^{m} \Psi_{(mb)} +
{1 \over 3} g_{ba}  D^{m} \Psi_{(mc)} ) + i M \; \Psi_{a[bc]} = 0 \; ,
$$
$$
\eqno(4e)
$$
$$
{\lambda_{12} \over 3} \; \left [   D_{a} \Psi_{(bc)} +   D_{b}
\Psi_{(ca)} +
 D_{c} \Psi_{(ab)} - \right.
 $$
 $$
\left. - {1\over 3}\; g_{ac} D^{m}\Psi_{(mb)} - {1\over 3}\;
g_{cb} D^{m}\Psi_{(ma)} - {1\over 3}\; g_{ba} D^{m}\Psi_{(mc)}
\right ]  + iM \Psi_{(abc)}=0 \; .
$$
$$
\eqno(4f)
 $$

With the help of  (4e) and  (4f), let us express tensors  $\Psi_{a[bc]}$  and  $\Psi_{(abc)}$
through  the 2-rank tensor:
$$
  \Psi_{a[bc]}  = { i \lambda_{11} \over 2 M }\;
  \left (  \; D_{c} \Psi_{(ab)} - D_{b}
\Psi_{(ac)} -
 {1 \over 3} g_{ca} D^{m} \Psi_{(mb)} +
{1 \over 3} g_{ba}  D^{m} \Psi_{(mc)}  \; \right )  \; ,
\eqno(5a)
$$
$$
 \Psi_{(abc)} = { i \lambda_{12} \over 3 M }
  \left  (\;  D_{a} \Psi_{(bc)} + D_{b}
\Psi_{(ca)} +
 D_{c} \Psi_{(ab)}  \right.
 $$
 $$
\left.  - {1\over 3}\; g_{ac} D^{m}\Psi_{(mb)} - {1\over 3}\;
g_{cb} D^{m}\Psi_{(ma)} - {1\over 3}\; g_{ba} D^{m}\Psi_{(mc)} \;
\right  )\,.
 \eqno(5b)
 $$

\noindent
Substitution it into eq. $(4d')$, we get
$$
{1  \over 6 (  \lambda_{1} \lambda_{4} +\lambda_{2} \lambda_{6})   }
  \; ( D_{a} B_{b}  + D_{b} B_{a} -{1 \over 2} \; g_{ab} D^{c} B_{c} )
$$
$$
+ {i  \over  M} \;
\left  [ (  \lambda_{7} \lambda_{4} + \lambda_{8} \lambda_{6})  - {1 \over 3}   \right  ] \;
\left ( D_{a} D^{c} \Psi_{(cb)}  + D_{b} D^{c} \Psi_{(ca)}  -
{1 \over 2} \; g_{ab} D^{c} D^{n} \Psi_{(nc)}  \right )
$$
$$
+  i {  2 \lambda_{9}\lambda_{12} \over 3  M }   D^{c}
  \left  (   D_{a} \Psi_{(bc)} + D_{b}
\Psi_{(ca)} +
 D_{c} \Psi_{(ab)}
 \right.
 $$
 $$
 \left. -  {1\over 3}  g_{ac} D^{m}\Psi_{(mb)} - {1\over 3}
g_{cb} D^{m}\Psi_{(ma)} - {1\over 3} g_{ba} D^{m}\Psi_{(mc)}
\right  )
$$
$$
-
i  {   \lambda_{10} \lambda_{11} \over  M }
\left [   D^{c}
\left (   D_{c} \Psi_{(ab)} - D_{b}
\Psi_{(ac)} -
 {1 \over 3} g_{ca} D^{m} \Psi_{(mb)} +
{1 \over 3} g_{ba}  D^{m} \Psi_{(mc)}   \right ) \right.
  $$
 $$
 \left. + D^{c}
 \left (   D_{c} \Psi_{(ba)} - D_{a}
\Psi_{(bc)} -
 {1 \over 3} g_{cb} D^{m} \Psi_{(ma)} +
{1 \over 3} g_{ab}  D^{m} \Psi_{(mc)}   \right ) \right ] +
  i M\; \Psi_{(ab)} =0 \; .
$$

Now, allowing for  (see (2a))
$$
\lambda_{10} \lambda_{11}  = {1  \over 2}  + {1 \over 3} \lambda_{9} \lambda_{12} \;, \qquad
\lambda_{4}\lambda_{7}  + \lambda_{6}\lambda_{8}   - {1 \over 3} = -  {8\over 9} \lambda_{9}\lambda_{12} \; .
$$

\noindent and using the notation $\lambda_{9}\lambda_{12} = \mu $, we obtain
$$
{M  \over 6 i(  \lambda_{1} \lambda_{4} +\lambda_{2} \lambda_{6})   }
  \left  ( D_{a} B_{b}  + D_{b} B_{a} -{1 \over 2} \; g_{ab} D^{c} B_{c}  \right )
$$
$$
-\;
   \mu   {8\over 9} D_{a} D^{c} \Psi_{(cb)}  -  \mu  {8\over 9} D_{b} D^{c} \Psi_{(ca)}  +
 \mu  {4\over 9}  \; g_{ab} D^{c} D^{n} \Psi_{(nc)}
$$
$$
+       \mu   {2 \over 3} D^{c}  D_{a} \Psi_{(bc)} +
 \mu  {2 \over 3}  D^{c} D_{b} \Psi_{(ca)} +
 \mu  {2 \over 3} D^{c} D_{c} \Psi_{(ab)} -
 \mu  {2\over 9}  g_{ac}  D^{c} D^{m}\Psi_{(mb)}
 $$
 $$
 - \;
  \mu {2\over 9} g_{cb}  D^{c} D^{m}\Psi_{(ma)} -
 \mu  {2\over 9} g_{ba}  D^{c} D^{m}\Psi_{(mc)}
  $$
$$
-   \;     {1  \over 2}   D^{c} D_{c} \Psi_{(ab)} +
{1  \over 2} D^{c} D_{b} \Psi_{(ac)} +
 {1 \over 6} g_{ca} D^{c} D^{m} \Psi_{(mb)} -
{1 \over 6} g_{ba}  D^{c}  D^{m} \Psi_{(mc)}
  $$
 $$
  -\;
   {1  \over 2}  D^{c}  D_{c} \Psi_{(ba)} +
 {1  \over 2}  D^{c} D_{a} \Psi_{(bc)} +
 {1 \over 6} g_{cb}  D^{c} D^{m} \Psi_{(ma)} -
 {1 \over 6} g_{ab}  D^{c} D^{m} \Psi_{(mc)}    \;   -
$$
$$
- \;
 \mu  {1 \over 3}   D^{c} D_{c} \Psi_{(ab)} +
\mu {1 \over 3} D^{c} D_{b} \Psi_{(ac)} +
 \mu {1 \over 9} g_{ca} D^{c} D^{m} \Psi_{(mb)} -
\mu {1 \over 9} g_{ba}  D^{c}  D^{m} \Psi_{(mc)}
 $$
 $$
  - \;
   \mu {1 \over 3}  D^{c}  D_{c} \Psi_{(ba)} +
  \mu {1 \over 3} D^{c} D_{a} \Psi_{(bc)} +
 \mu {1 \over 9} g_{cb}  D^{c} D^{m} \Psi_{(ma)} -
 \mu {1 \over 9} g_{ab}  D^{c} D^{m} \Psi_{(mc)}
$$
$$
 + \;
   M^{2} \; \Psi_{(ab)} =0 \; .
\eqno(4d'')
$$

\noindent
From  whence, after simple manipulations, we arrive at
(commutator will be noted as $ [ \hspace{2mm} , \hspace{2mm} ]_{-}$)
$$
{1  \over 6 i \; (  \lambda_{1} \lambda_{4} +\lambda_{2} \lambda_{6})   }
  \;  \left ( D_{a} B_{b}  + D_{b} B_{a} -{1 \over 2} \; g_{ab} D^{c} B_{c}  \right )
$$
$$
 - {1 \over M} \left [  D^{c}  D_{c} \Psi_{(ba)}
 - {1  \over 2}  \;( \;  D^{c} D_{b} \Psi_{(ac)}  +   D^{c} D_{a} \Psi_{(bc)} \; )\;
  \right.
 $$
 $$
 \left. + {1 \over 3} \; g_{ab} \;  D^{n}D^{m} \Psi_{(nm)}-
{1 \over 6}  \; ( \;  D_{a} D^{m} \Psi_{(mb)}   + D_{b} D^{m} \Psi_{(ma)} \; ) \right ]
$$
$$
+ \; { \mu \over  M}  \left (  [D^{c}, D_{a}]_{-} \Psi_{(bc)}  +  [D^{c}, D_{b}]_{-} \Psi_{(ac)}  \right  )
 +
   M  \; \Psi_{(ab)} =0 \; .
$$
$$
\eqno(4d''')
$$

Now, let us introduce a new variable   (constant  $\gamma$  will be specified below)
$$
\Phi_{[bc]a} = - {1 \over M}{ \gamma \over 2}  \left (  D_{c} \Psi_{(ab)} - D_{b} \Psi_{(ac)}  +
 {1\over 3} g_{ab} D^{m} \Psi_{(mc)}  - {1\over 3} g_{ac} D^{m} \Psi_{(mb)} \right ) ,
$$

\noindent
then we derive an identity
$$
{1 \over  \gamma } \; \left ( D^{c} \Phi_{[bc]a} +  D^{c} \Phi_{[ac]b}
\right )  = -{ 1  \over M}
$$
$$
\times
\left [
{1 \over 2} \left (  D^{c} D_{c} \Psi_{(ab)} -  D^{c} D_{b} \Psi_{(ac)}  +
 {g_{ab}\over 3}  D^{c} D^{m} \Psi_{(mc)}  - {g_{ac}\over 3}   D^{c} D^{m} \Psi_{(mb)} \right )
 \right.
$$
$$
\left.
+
 {1 \over 2} \left (  D^{c}  D_{c} \Psi_{(ba)} -  D^{c} D_{a} \Psi_{(bc)}  +
 {g_{ba} \over 3}   D^{c} D^{m} \Psi_{(mc)}  - {g_{bc}\over 3}   D^{c} D^{m} \Psi_{(ma)} \right )
\right ]
$$
$$
=  -{ 1 \over M}   \left  (   D^{c} D_{c} \Psi_{(ab)} -  {1 \over 2} D^{c} D_{b} \Psi_{(ac)}
- {1 \over 3}  D^{c} D_{a} \Psi_{(bc)} + {g_{ab} \over 3}  D^{c} D^{m} \Psi_{(mc)}
 \right.
$$
$$
\left. -
{g_{ac}\over 6}   D^{c} D^{m} \Psi_{(mb)}
- {g_{bc}\over 6 }   D^{c} D^{m} \Psi_{(ma)} \right )
$$

\noindent which coincides with the expression in rackets in
$(4d''')$. Therefore, eq.  $(4d''')$ may be presented as  (let it be
$\gamma = \sqrt{2}$)
$$
{1  \over 6 i \; (  \lambda_{1} \lambda_{4} +\lambda_{2} \lambda_{6})   }
  \;  \left ( D_{a} B_{b}  + D_{b} B_{a} -{1 \over 2} \; g_{ab} D^{c} B_{c}  \right ) \;
  +   {1 \over  \sqrt{2} }  \; \left ( D^{c} \Phi_{[bc]a} +  D^{c} \Phi_{[ac]b}
\right )
 $$
$$
+ \; { \mu \over  M}  \left ( \;  [D^{c}, D_{a}]_{-} \Psi_{(bc)}  +  [D^{c}, D_{b}]_{-} \Psi_{(ac)} \; \right  )
 +
   M  \; \Psi_{(ab)} =0 \;.
$$
$$
\eqno(4d'''')
$$

In the following, it will be convenient to use  two variables
$$
\Phi =-\; {1 \over  4 \sqrt{3} (  \lambda_{1} \lambda_{4} +\lambda_{2} \lambda_{6})   } \; \Psi \;,
\qquad
\Phi_{a} ={i \over   \sqrt{6} (  \lambda_{1} \lambda_{4} +\lambda_{2} \lambda_{6})   } \; B_{a}
\;.
$$

Thus, from  50-component system, we have arrives  at
a modified  30-component  model
$$
{1 \over \sqrt{2}}  D^{a} \Phi _{a} +  M\; \Phi = 0 \; ,
$$
$$
 {1 \over \sqrt{2}} \;  D_{a} \Phi + {\sqrt {2\over 3}}  \; D^{b} \Psi_{(ba)} +   M\; \Phi_{a} =0 \; ,
$$
$$
-{1  \over \sqrt{6}   }
  \;   ( \;  D_{a} \Phi_{b}  + D_{b} \Phi_{a} -{1 \over 2} \; g_{ab} D^{c} \Phi_{c}  \;  ) \;
  +   {1 \over  \sqrt{2} }  \;  (  \; D^{c} \Phi_{[bc]a} +  D^{c} \Phi_{[ac]b}
\;  )
 $$
$$
+ \; { \mu \over  M}  \; ( \;  [D^{c}, D_{a}]_{-} \Psi_{(bc)}  +  [D^{c}, D_{b}]_{-} \Psi_{(ac)} \;   )
 +
   M  \; \Psi_{(ab)} =0 \; ,
$$
$$
 { 1 \over \sqrt{2}}  \;  ( \;  D_{c} \Psi_{(ab)} - D_{b} \Psi_{(ac)}  +
 {1\over 3} g_{ab} D^{m} \Psi_{(lc)}  - {1\over 3} g_{ac} D^{m} \Psi_{(lb)} \;  ) + M \; \Phi_{a[bc]} = 0 \; .
$$

By the simple linear transformations
$$
\Phi  =  - \tilde{\Phi}  \; , \qquad \Psi_{a}  = \sqrt{2} \; \tilde{\Phi}_{a} \; , \qquad
\Phi_{(ab)} = {1 \over  \sqrt{3}} \; \tilde{\Phi}_{(ab)}, \qquad
\Phi_{[bc]a} =  {1 \over \sqrt{6}} \; \tilde{\Phi}_{[bc]a}
\eqno(5)
$$

\noindent it becomes simpler
$$
  D^{a} \tilde{\Phi} _{a} -  M\; \tilde{ \Phi} = 0 \; ,
$$
$$
 {1 \over 2} \;  D_{a} \tilde{\Phi} - {1 \over 3}   \; D^{b} \tilde{\Psi}_{(ba)}
 -   M\;  \tilde{\Phi}_{a} =0 \; ,
$$
$$
  ( \;  D_{a} \tilde{\Phi}_{b}  + D_{b} \tilde{\Phi}_{a} -{1 \over 2} \; g_{ab} D^{c} \tilde{\Phi}_{c}  \;  ) \;
  +   {1 \over  2 }  \;  (  \;  D^{c} \tilde{\Phi}_{[ca]b} + D^{c} \tilde{\Phi}_{[cb]a}
\;  )
 $$
$$
- \;  {\mu \over    M}  \;  \left ( \;  [D^{c}, D_{a}]_{-} \; \tilde{\Phi}_{(bc)}  +
[D^{c}, D_{b}]_{-}  \; \tilde{\Phi}_{(ac)} \;   \right )
 -
   M  \; \tilde{\Phi}_{(ab)} =0 \; ,
$$
$$
 D_{c} \tilde{\Phi}_{(ba)} - D_{b} \tilde{\Phi}_{(ca)}  +
 {1\over 3} g_{ba} D^{m} \tilde{\Psi}_{(mc)}  - {1\over 3} g_{ca} D^{m} \tilde{\Psi}_{(mb)}
 - M \;  \tilde{\Phi}_{[cb]a} = 0 \; .
$$
$$
\eqno(7)
$$

\noindent
If  $\mu =0$, we obtain a  30-component theory (the sign of  $\sim$ is taken away):
$$
D^{a} \Phi _{a} - M \; \Phi  =0 \; ,
$$
$$
{1 \over 2} \; D_{a} \Phi  -  {1 \over 3} \;
D^{b} \Phi_{(ab)} - M \; \Phi_{a}  = 0 \; ,
$$
$$
D_{a} \Phi_{b} \; +\;  D_{b} \Phi_{a} -
{1 \over 2} g_{ab} \;  D^{c} \Phi_{c}
+{1 \over 2} \; ( D^{c} \Phi_{[ca]b}   +  D^{c}
\Phi_{[cb]a} \; )  -  M \; \Phi_{(ab)}  = 0\;
,
$$
$$
 D_{c} \Phi_{(ba)} - D_{b} \Phi_{(ca)}  +
 {1\over 3} g_{ba} D^{m} \Phi_{(mc)}  - {1\over 3} g_{ca} D^{m} \Phi_{(mb)}
 - M \;  \Phi_{[cb]a} = 0 \; .
$$
$$
\eqno(8)
$$

Comparing it with (3), we note differences in 3-d equation. However it is easily demonstrated their equivalence.
Indeed, from 4-th equation we derive
$$
g^{ac}\,D_{a}\,\Phi_{(bc)}-g^{ac}\,D_{b}\,\Phi_{(ac)}+{1\over 3}(g^{ac}\,g_{bc}\,D^{k}\,\Phi_{(ak)}-g^{ac}\,g_{ac}\,D^{k}\,\Phi_{(bk)})-M\,g^{ac}\,\Phi_{[ab]c}=0\,.
$$
\noindent Because  $g^{ac}\,\Phi_{(ac)}=0,\; g^{ac}\,g_{ac}=4$, then
$$
\Phi^{b}_{[ab]}=0\,,\eqno(9)
$$

\noindent which means that the term ${1\over 2}\,g_{ab}\,D^{c}\,\Phi^{n}_{[cn]}$
in 3-d equation in  (3) vanishes identically. Thus, systems (3) and (8) are  equivalent.

Let us find an explicit form for  (see 4-th eq. in (7))
$$
\mu   M^{-1}  \;  \left ( \;  [D^{c}, D_{a}]_{-} \Phi_{(bc)}  +
[D^{c}, D_{b}]_{-} \Phi_{(ac)} \;   \right ).
\eqno(10)
$$

\noindent It suffices to consider the first term
$$
[D^{c}, D_{a}]_{-} \tilde{\Phi}_{(bc)}  =
  [ \nabla_{c} + ieA_{c},  \nabla_{a} + ieA_{a} ]_{-} \Phi_{b}^{\;\;c}
$$
$$
= (\nabla_{c}  \nabla_{a}  - \nabla_{a}  \nabla_{c} )\;  \Phi_{b}^{\;\;c} + ie F_{ca}\;  \Phi_{b}^{\;\;c}.
$$

\noindent with the help of known rules
$$
( \nabla_{c} \nabla_{a} -   \nabla_{a} \nabla_{c} )\; A_{bk} =
- A_{nk} \; R^{n}_{\;\;b\; ca} -  A_{bn} \; R^{n}_{\;\;k\;ca} \; .
$$

\noindent
from  whence it follows
$$
( \nabla_{c} \nabla_{a} -   \nabla_{a} \nabla_{c} )\; A_{b}^{\;\;c} =
- A_{n}^{\;\;c} \; R^{n}_{\;\;b\; ca} -  A_{bn} \; R^{nc}_{\;\;\;\;\;ca}\; .
$$

\noindent Further, allowing for symmetry of curvature tensor we find
$$
( \nabla^{c} \nabla_{a} -   \nabla_{a} \nabla^{c} )\; A_{bc} =
  R_{ca\;bn} \; A^{nc}  +     \; A_{b}^{\;\;n}\; R_{na} \; ,
$$

\noindent  we derive
$$
( \nabla^{c} \nabla_{a} -   \nabla_{a} \nabla^{c} )\; \Phi _{bc} =
  R_{ca\;bn} \; \Phi^{cn}  +      R_{ac}\; \Phi_{\;\;b}^{c} \; .
\eqno(12)
$$

\noindent  Therefore,
$$
[D^{c}, D_{a}]_{-} \tilde{\Phi}_{(bc)}  = ie F_{ca} \Phi_{b}^{\;\;c} +
R_{ca\;bn} \; \Phi^{cn}  +      R_{ac}\; \Phi_{\;\;b}^{c} \; ,
\eqno(13)
$$

\noindent and additional  interaction term
is specified by
$$
{\mu   \over  M }  \; \left ( \;  [D^{c}, D_{a}]_{-} \; \Phi_{(bc)}  +
[D^{c}, D_{b}]_{-}  \; \Phi_{(ac)} \;   \right )
$$
$$
=
{\mu \over   M }  \;  ie  \; (   \Phi_{a}^{\;\;c}  F_{cb}  +   \Phi_{b}^{\;\;c}  F_{ca} ) +
{\mu \over   M } (\;  R_{ca\;bn} \; \Phi^{cn}  + R_{cb\;an} \; \Phi^{cn}   \;  )
$$
$$
+
 {\mu \over   M }  ( \;  R_{ac}\; \Phi_{\;\;b}^{c}   + R_{bc}\; \Phi_{\;\;a}^{c} \;   ) \; .
\eqno(14)
$$

Relation  (14) means that the parameter $\mu$, initially interpreted as defining anomalous magnetic moment,
also determines additional interaction with geometrical background,
through Ricci  $R_{kl}$ and Riemann curvature tensor $R_{klmn}$.

It should be noted that in the case of spin  1/2 particle with anomalous magnetic  moment
 \cite{1955-Petras(1), 1955-Petras(2), 1955-Formanek, 1979-Bogush-Kisel,
 1984-Bogush(2)} there   arises additional interection through Ricci scalar \cite{A-12, A-60};
in the case of spin 1 particle  \cite{1969-Capri(1), 1969-Capri(2), 1971-Shamali, 1972-Capri, 1973-Shamaly,
1974-Hagen} there  arises an additional interaction through  Ricci tensor \cite{A-65}.
In other words,  sensitiveness of the anomalous magnetic  moment  to the space-time geometry
substantially depends on spin of the particle.

Authors are grateful to  participants of seminar  of Laboratory of theoretical physics,
Institute of Physics of National Academy of Sciences of Belarus for discussion and advices.

\end{document}